\documentstyle[12pt]{article}
\begin{document}\bibliographystyle{plain}
\begin{titlepage}\renewcommand{\thefootnote}{\fnsymbol{footnote}}
\hfill\begin{tabular}{l}HEPHY-PUB
693/98\\UWThPh-1998-38\\hep-ph/9904391\\April 1999\end{tabular}\\[3cm]
\huge\begin{center}{\bf QUALITY OF VARIATIONAL\\TRIAL STATES}\\
\vspace{2cm}\large{\bf Wolfgang LUCHA\footnote[1]{\normalsize\ {\em
E-mail\/}: wolfgang.lucha@oeaw.ac.at}}\\[.5cm]Institut f\"ur
Hochenergiephysik,\\\"Osterreichische Akademie der
Wissenschaften,\\Austria\\[2cm]{\bf Franz F.
SCH\"OBERL\footnote[2]{\normalsize\ {\em E-mail\/}:
franz.schoeberl@univie.ac.at}}\\[.5cm]Institut f\"ur Theoretische
Physik,\\Universit\"at Wien, Austria\vfill{\normalsize\bf
Abstract}\end{center}\small Various measures for the accuracy of approximate
eigenstates of semibounded self-adjoint operators $H$ in quantum theory,
derived, e.g., by some variational technique, are scrutinized. In particular,
the matrix elements of the commutator of the operator $H$ and (suitably
chosen) different operators with respect to degenerate approximate
eigenstates of $H$ obtained~by~the variational methods are proposed as new
criteria for the accuracy of variational~eigenstates.

\vspace{3ex}

\renewcommand{\thefootnote}{\arabic{footnote}}\end{titlepage}

\normalsize

\section{Motivation}A central element in quantum theory is the solution of
eigenvalue problems. However, usually there is no suitable exact solution to
which perturbation theory can~be~applied.

A very efficient way to locate the discrete spectrum of some self-adjoint
operator~$H$ bounded from below is provided by the famous Rayleigh--Ritz
variational technique~\cite{MMP}: {\sl If the eigenvalues $E_k$,
$k=0,1,\dots$, of $H$ are ordered according to $E_0\le E_1\le E_2\le\dots$,
the first $d$ of them are bounded from above by the $d$ eigenvalues $\widehat
E_k$, $k=0,1,\dots,d-1$, (ordered by $\widehat E_0\le\widehat
E_1\le\dots\le\widehat E_{d-1}$) of that operator which is obtained by
restricting~$H$ to some $d$-dimensional subspace of the domain of $H$, i.e.,
$E_k\le\widehat E_k$, $k=0,1,\dots,d-1$.} However, frequently it is not
straightforward~to quantify how close approximate and exact eigenstates are.
Thus, we embark upon a systematic study of the accuracy of~the variationally
determined eigenstates of $H$ and suitable measures to judge their quality.

\section{Measures of the Quality of Trial States}\label{Sec:MQ}Consider some
self-adjoint operator $H$, $H^\dagger=H$, assumed to be bounded from below.
Suppressing, for the moment, the index $k=0,1,2,\dots,$ let the eigenvalue
equation for $H$,\begin{equation}H|\chi\rangle=E|\chi\rangle\
,\label{eq:GEVE}\end{equation}be solved by some (generic) eigenvector
$|\chi\rangle$ corresponding to some (real) eigenvalue $E$. The
Rayleigh--Ritz variational technique yields an upper bound $\widehat E$ on
this eigenvalue~$E$ as well as, by diagonalization of the relevant
characteristic equation, the corresponding vector $|\varphi\rangle$ in the
$d$-dimensional trial space. There exist several (potentially meaningful)
measures~of the quality of this trial state $|\varphi\rangle$ which
immediately come to one's mind:\begin{enumerate}\item The trial state
$|\varphi\rangle$ is supposed to represent---to a certain degree of
accuracy---the approximate solution of the eigenvalue problem defined in
Eq.~(\ref{eq:GEVE}). Consequently, a first indicator for the resemblance of
$|\varphi\rangle$ with the exact eigenstate $|\chi\rangle$ would~be the
distance between the expectation value of the operator $H$ with respect
to~the trial state $|\varphi\rangle$, i.e., between the obtained upper bound
$\widehat E\equiv
\langle\varphi|H|\varphi\rangle/\langle\varphi|\varphi\rangle$, and the exact
eigenvalue $E$. However, the precise location of the exact eigenvalue~$E$~is
usually not known.\item The natural measure for the resemblance of the
Hilbert-space vectors $|\varphi\rangle$ and~$|\chi\rangle$ under
consideration is the overlap
\begin{equation}S\equiv\frac{\langle\varphi|\chi\rangle}
{\sqrt{\langle\varphi|\varphi\rangle\,\langle\chi|\chi\rangle}}\label{eq:OTE}
\end{equation}of the trial state $|\varphi\rangle$ with the eigenstate
$|\chi\rangle$.\item Consider the commutator $[G,H]$ of the operator $H$
under consideration with~any other operator $G$, where the domain of $G$ is
assumed to contain the domain~of~$H$. Then the expectation value of this
commutator with respect to a given eigenstate $|\chi\rangle$ of $H$
vanishes:\begin{equation}\langle\chi|[G,H]|\chi\rangle=0\
.\label{eq:VC}\end{equation}Hence, choosing different operators $G$ generates
a whole class of operators $[G,H]$ each of which may serve to test the
quality of a given trial state $|\varphi\rangle$ by evaluating how close the
expectation value $\langle\varphi|[G,H]|\varphi\rangle$ with respect to
$|\varphi\rangle$ comes to zero. This expectation value vanishes, of course,
also if, by accident, the state $|\varphi\rangle$~is~an eigenstate of $G$.
However, for a given operator $G$, after having determined
$|\varphi\rangle$,~it is straightforward to check for this circumstance, for
instance, by inspecting the variance of $G$ with respect to
$|\varphi\rangle$; the latter vanishes if $|\varphi\rangle$ is an
eigenstate~of~$G$. Moreover, it goes without saying that an expectation value
$\langle\varphi|[G,H]|\varphi\rangle$ vanishes also if the state
$|\varphi\rangle$ is an eigenstate of the commutator $[G,H]$ with
eigenvalue~0, or even if the state defined by $[G,H]|\varphi\rangle$ proves
to be orthogonal to the state~$|\varphi\rangle$.

For any self-adjoint operator $G$, i.e., $G^\dagger=G$, this commutator is
anti-Hermitean, which clearly suggests to define a self-adjoint operator
$C=C^\dagger$ (on the domain of $H$) by $[G,H]=:{\rm i}\,C$. If, for example,
$G$ is chosen to be the
generator~of~dilations,\begin{equation}G\equiv\frac{1}{2}\,({\bf x}\cdot{\bf
p}+{\bf p}\cdot{\bf x})\ ,\label{eq:DG}\end{equation}the relation
(\ref{eq:VC}) is precisely the ``master virial theorem'' introduced in
Ref.~\cite{Lucha90:RVTs}~for a systematic study of (relativistic) virial
theorems \cite{Lucha89:RVT}. In this case, for operators $H$ of the form of
some typical Hamiltonian consisting of a momentum-dependent kinetic-energy
operator, $T({\bf p})$, and a coordinate-dependent interaction-potential
operator, $V({\bf x})$, that is, $H=T({\bf p})+V({\bf x})$, the operator $C$
becomes the ``virial operator''\begin{equation}C={\bf
p}\cdot\frac{\partial}{\partial{\bf p}}T({\bf p})-{\bf
x}\cdot\frac{\partial}{\partial{\bf x}}V({\bf x})\
.\label{eq:VO}\end{equation}The point spectrum (i.e., the set of all
eigenvalues) of the dilation generator~(\ref{eq:DG}) is empty; in other
words, the dilation generator has no Hilbert-space eigenvectors.
\end{enumerate}

\section{Spinless Salpeter Equation}\label{Sec:SSE}Let us apply the above
general considerations to the prototype of all (semi-) relativistic
bound-state equations, the ``spinless Salpeter equation,'' defined by the
Hamiltonian (in one-particle form, encompassing also the equal-mass
two-particle case \cite{Lucha94varbound,Lucha96:AUB,Lucha98O,Lucha98D})
\begin{equation}H=T+V\ ;\label{Eq:SRH}\end{equation}here $T$ is the
relativistic kinetic energy of some particle of mass $m$ and momentum~${\bf
p}$,$$T=T({\bf p})\equiv\sqrt{{\bf p}^2+m^2}\ ,$$and $V=V({\bf x})$ is an
arbitrary, coordinate-dependent, static interaction potential. The spinless
Salpeter equation is then just the eigenvalue equation for $H$,
$H|\chi_k\rangle=E_k|\chi_k\rangle,$ $k=0,1,2,\dots,$ for the set of
eigenvectors $|\chi_k\rangle$ corresponding to energy eigenvalues $E_k$.
Analytic upper bounds $\widehat E_k$ on these eigenvalues have been given
\cite{Lucha94varbound,Lucha96:AUB,Lucha98O,Lucha98D,Lucha96:CCC,Lucha97:L,
Lucha98R}.

For the sake of comparison, we focus our interest to central potentials
$V({\bf x})=V(r)$, $r\equiv|{\bf x}|$. Furthermore, in order to facilitate
the numerical treatment of the problem,~we consider the harmonic-oscillator
potential\begin{equation}V(r)=a\,r^2\ ,\quad a>0\
.\label{eq:HOP}\end{equation}The reason for this particular choice is the
following: In momentum space, the operator $r^2$ is represented by the
Laplacian with respect to the momentum~${\bf p}$, $r^2\rightarrow-\Delta_{\bf
p}$, while the kinetic energy $T$, nonlocal in configuration space, is
represented by a multiplication operator. Consequently, exactly for a
harmonic-oscillator potential the semirelativistic Hamiltonian $H$ in its
momentum-space representation is equivalent to a nonrelativistic Hamiltonian
with some (effective) interaction potential reminiscent of the square
root:\begin{equation}H=-a\,\Delta_{\bf p}+\sqrt{{\bf p}^2+m^2}\
.\label{eq:NRH}\end{equation}The solutions of the corresponding eigenvalue
equation may then be found with one of the numerous procedures designed for
the treatment of the nonrelativistic Schr\"odinger equation.

For the harmonic-oscillator potential, it is comparatively easy to get a
first idea~of the approximate location of the energy levels $E_k$ by entirely
analytical considerations:\begin{itemize}\item On the one hand, every
eigenvalue $E_k$ is bounded from above by the eigenvalue $E_{k,{\rm NR}}$ of
the nonrelativistic counterpart of $H$: $E_k\le E_{k,{\rm NR}}$.\item On the
other hand, every eigenvalue $E_k$ is bounded from below by the eigenvalue
$E_k(m=0)$ of the Hamiltonian $H$ corresponding to a vanishing particle mass
$m$: $E_k\ge E_k(m=0)$.\end{itemize}

\section{The ``Laguerre'' Trial Space}As far as the achieved accuracy of the
solutions obtained is concerned, the most crucial step in all variational
games of the Rayleigh--Ritz kind is, for a given operator $H$ under
consideration, a reasonable definition of the adopted trial subspace of
the~domain~of~$H$.

For spherically symmetric potentials $V(r)$, a very popular choice for the
basis states which span the trial space required for the application of the
variational technique are ``Laguerre'' trial states, defined in
configuration-space representation by \cite{LTS,Lucha97:L,Lucha98O,Lucha98D}
\begin{equation}\psi_{k,\ell m}({\bf
x})=\sqrt{\frac{(2\,\mu)^{2\,\ell+2\,\beta+1}\,k!}
{\Gamma(2\,\ell+2\,\beta+k+1)}}\,r^{\ell+\beta-1}\exp(-\mu\,r)\,
L_k^{(2\,\ell+2\,\beta)}(2\,\mu\,r)\,{\cal Y}_{\ell m}(\Omega_{\bf x})\
,\label{eq:LTF}\end{equation}where $L_k^{(\gamma)}(x)$ denote the generalized
Laguerre polynomials (for the parameter $\gamma$) \cite{Abramowitz} and
${\cal Y}_{\ell m}(\Omega)$ are the spherical harmonics for angular momentum
$\ell$ and its projection~$m$. The trial functions (\ref{eq:LTF}) involve two
variational parameters, $\mu$ (with dimension of~mass) and $\beta$
(dimensionless), which, by the requirement of normalizability of these
functions, are subject to the constraints $\mu>0$ and $2\,\beta>-1$.

One of the advantages of the trial function (\ref{eq:LTF}) is the easy
availability of an~analytic expression for the corresponding momentum-space
representation of these trial states.

For the present investigation, we too employ the ``Laguerre'' trial states
defined by Eq.~(\ref{eq:LTF}), with, for both definiteness and ease of
calculation, the variational parameters $\mu$ and $\beta$ kept fixed to the
values $\mu=m$ and $\beta=1$.

\section{Rates of Convergence of the Quality Measures}Now, let us observe our
variational eigenstates, $|\varphi\rangle$, approaching the exact
eigenstates, $|\chi\rangle$, for increasing dimension $d$ of the employed
trial space, by comparing the~behaviour of the various measures for the
accuracy of approximate eigenstates discussed in Sec.~\ref{Sec:MQ}.

Without doubt, the only genuine ``point of reference'' of any variational
solution~to an eigenvalue problem is the corresponding exact solution. The
exact solution sought~is computed here with the help of the numerical
integration procedure developed for the solution of the nonrelativistic
Schr\"odinger equation in Ref.~\cite{Falkenst85}.

Table~\ref{Tab:QVS} confronts, for the ground state and the lowest radial and
orbital excitations, the approximate solutions as calculated with the help of
the Rayleigh--Ritz variational technique for ``Laguerre'' trial subspaces of
the domain of $H$ of increasing dimension~$d$ with the exact solutions of the
eigenvalue problem for the semirelativistic Hamiltonian (\ref{Eq:SRH}) with a
central interaction potential of the harmonic-oscillator form (\ref{eq:HOP}).
First of~all, as discussed in Sec.~\ref{Sec:SSE}, the exact position of any
eigenvalue $E$ of our Hamiltonian $H$ is confined to a range defined by the
nonrelativistic upper bound $E_{\rm NR}$ and the~zero-mass lower bound
$E(m=0)$ on this energy eigenvalue $E$. There are several quantities~which
may participate in a competition for ``the best or most reasonable measure of
quality:''\begin{enumerate}\item The relative error
$\varepsilon\equiv(\widehat E-E)/E$ of every upper bound $\widehat E$ on the
exact energy eigenvalue $E$ is, by definition, always nonnegative, i.e.,
$\varepsilon\ge0$.

\small\begin{table}[ht]\caption[]{\small Characterization of the quality of
the variational solution of the eigenvalue problem of the semirelativistic
Hamiltonian $H=\sqrt{{\bf p}^2+m^2}+V(r)$ with harmonic-oscillator potential
$V(r)=a\,r^2$, for states of radial quantum number $n_{\rm r}=0,1,2$ and
orbital angular momentum $\ell=0,1,2$ (called 1S, 2S, 3S, 1P, and 1D in usual
spectroscopic notation), obtained with the help of our ``Laguerre'' trial
states spanning trial spaces of increasing dimension $d=1,2,25$, by: the
nonrelativistic upper bound $E_{\rm NR}$ and zero-mass lower bound $E(m=0)$
on the energy, the (numerically computed) ``exact'' energy $E$, the
variational upper bound $\widehat E$ on this energy, the relative error
$\varepsilon$ of the upper bound, the deviation from unity, $\sigma$, of the
overlap squared~of exact and variational eigenstates, the (appropriately
normalized) expectation values $\nu$ of the virial operator $C$, and the
(normalized) maximum local difference $\omega$ of the momentum-space
representations of exact and variational eigenstates. The physical parameters
are fixed to~the values $m=2\;\mbox{GeV}$ for the particle mass and
$a=2\;\mbox{GeV}^3$ for the harmonic-oscillator coupling. A simple entry
``0'' indicates that the numerical value is closer to 0 than the rounding
error.}\label{Tab:QVS}\footnotesize
\begin{center}\begin{tabular}{lrlllll}\hline\hline&&\\[-1.5ex]
Quantity&\multicolumn{1}{c}{$d$}&\multicolumn{5}{c}{State}\\[1ex]
\cline{3-7}\\[-1.5ex]&&\multicolumn{1}{c}{1S}&\multicolumn{1}{c}{2S}&
\multicolumn{1}{c}{3S}&\multicolumn{1}{c}{1P}&\multicolumn{1}{c}{1D}\\[1ex]
$n_{\rm r}$&&\multicolumn{1}{c}{0}&\multicolumn{1}{c}{1}&
\multicolumn{1}{c}{2}&\multicolumn{1}{c}{0}&\multicolumn{1}{c}{0}\\
$\ell$&&\multicolumn{1}{c}{0}&\multicolumn{1}{c}{0}&\multicolumn{1}{c}{0}&
\multicolumn{1}{c}{1}&\multicolumn{1}{c}{2}\\[1ex]\hline&&\\[-1.5ex]$E_{\rm
NR}$ [GeV]&&4.12132&6.94975&9.77817&5.53553&6.94975\\[1ex]$E(m=0)$
[GeV]&&2.94583&5.15049&6.95547&4.23492&5.35234\\[1ex]
\hline&&\\[-1.5ex]$E$
[GeV]&&3.82493&5.79102&7.48208&4.90145&5.89675\\[1ex]\hline&&\\[-1.5ex]
$\widehat E$ [GeV]&1&4.21624&---&---&6.50936&9.77866\\
&2&3.92759&8.10850&---&5.24154&7.18242\\
&25&3.82494&5.79114&7.48290&4.90149&5.89681\\[1ex]\hline&&\\[-1.5ex]
$\varepsilon$&1&0.1023&---&---&0.3280&0.6583\\
&2&0.0268&0.4002&---&0.0694&0.2180\\
&25&0&0&0.0001&0&0\\[1ex]\hline&&\\[-1.5ex]$\sigma$
&1&0.09618&---&---&0.36144&0.65587\\ &2&0.02375&0.43693&---&0.09001&0.34398\\
&25&0&0&0.00008&0&0\\[1ex]\hline&&\\[-1.5ex]$\nu$
&1&$-0.6120$&---&---&$-0.8328$&$-0.9074$\\
&2&$+0.0308$&$-0.8666$&---&$-0.5103$&$-0.7483$\\
&25&0&$-0.0001$&$+0.0001$&0&0\\[1ex]\hline&&\\[-1.5ex]$\omega$
&1&$+0.9277$&---&---&$+0.7541$&$+$1.0578\\
&2&$-0.00754$&$+2.4577$&---&$+0.3598$&$+0.7262$\\
&25&$+0.00003$&$-0.0017$&$+0.0002$&$+0.0004$&$+0.0003$\\[1ex]
\hline\hline\end{tabular}\end{center}\end{table}\normalsize
\item The deviation from unity, $\sigma$, of the modulus squared of the
overlap $S$ of exact~and variational eigenstates defined in
Eq.~(\ref{eq:OTE}), $\sigma\equiv1-|S|^2\label{eq:1-Oý}$, is clearly confined
to the range $0\le\sigma\le1$.\item The use of the expectation values of the
commutators $[G,H]$ with respect to~the variational eigenstates
$|\varphi\rangle$ is illustrated for the particular example of the dilation
generator $G$ defined in Eq.~(\ref{eq:DG}), by considering (suitably
normalized) expectation values $\langle\varphi|C|\varphi\rangle$ of the
virial operator $C$ given in Eq.~(\ref{eq:VO}):
$$\nu\equiv\frac{\langle\varphi|C|\varphi\rangle} {\langle\varphi|{\bf
x}\cdot\frac{\partial}{\partial{\bf x}}V({\bf x})
|\varphi\rangle}=\frac{\langle\varphi|{\bf p}\cdot\frac{\partial}
{\partial{\bf p}}T({\bf p})|\varphi\rangle}{\langle\varphi|{\bf
x}\cdot\frac{\partial}{\partial{\bf x}}V({\bf x}) |\varphi\rangle}-1\
.$$\item Finally, the normalized maximum difference of the normalized
momentum-space representations $\tilde\varphi({\bf p})$ and $\tilde\chi({\bf
p})$ of variational eigenstate $|\varphi\rangle$ and exact eigenstate
$|\chi\rangle$, respectively, i.e., the maximum pointwise relative error in
momentum space, $\omega\equiv\max_{\bf p}[\tilde\varphi({\bf
p})-\tilde\chi({\bf p})]/\max_{\bf p}\tilde\chi({\bf p})$ is
listed.\end{enumerate}Note that the only measure for the accuracy of
approximate eigenstates $|\varphi\rangle$ which does not require any
information other than the one provided by the variational technique is
$\nu$, the (normalized) expectation values of the commutator $[G,H]$ with
respect to~$|\varphi\rangle$. Inspection of Table~\ref{Tab:QVS} reveals that
$\nu$ represents indeed a sensitive measure of quality:~for increasing
trial-space dimension $d$ it converges to zero at roughly the same rate as
both energy and overlap error, $\varepsilon$ and $\sigma$, but makes more
sense than a pointwise error like~$\omega$.

\section{Summary and Conclusions}Various measures for the accuracy of
approximate eigenstates of arbitrary (self-adjoint, semibounded) operators
$H$ have been studied. The vanishing of the expectation values of the
commutator of $H$ and any other well-defined operator, taken with respect
to~the approximate eigenstates, provides a useful set of criteria for
estimating the significance of the variational solution. This has been
illustrated by considering the commutator~of the Hamiltonian of the spinless
Salpeter equation and the generator of dilations.

\section*{Acknowledgements}We would like to thank H.~Narnhofer for
stimulating discussions and a critical reading of the manuscript.


\begin{thebibliography}{30}
\bibitem{MMP}M.~Reed and B.~Simon, {\em Methods of Modern Mathematical
Physics~IV: Analysis~of Operators\/} (Academic Press, New York, 1978)
Sec.~XIII.1 and XIII.2.
\bibitem{Lucha90:RVTs}W.~Lucha and F.~F.~Sch\"oberl, Mod.~Phys.~Lett. {\bf
A5} (1990) 2473.
\bibitem{Lucha89:RVT}W.~Lucha and F.~F.~Sch\"oberl, Phys.~Rev.~Lett. {\bf 64}
(1990) 2733.
\bibitem{Lucha94varbound}W.~Lucha and F.~F.~Sch\"oberl, Phys.~Rev.~D {\bf 50}
(1994) 5443.
\bibitem{Lucha96:AUB}W.~Lucha and F.~F.~Sch\"oberl, Phys.~Rev.~A {\bf 54}
(1996) 3790.
\bibitem{Lucha98O}W.~Lucha and F.~F.~Sch\"oberl, Int. J. Mod. Phys. A {\bf
14} (1999) 2309.
\bibitem{Lucha98D}W.~Lucha and F.~F.~Sch\"oberl, Fizika B {\bf 8} (1999) 193.
\bibitem{Lucha96:CCC}W.~Lucha and F.~F.~Sch\"oberl, Phys.~Lett.~B {\bf 387}
(1996) 573.
\bibitem{Lucha97:L}W.~Lucha and F.~F.~Sch\"oberl, Phys.~Rev.~A {\bf 56}
(1997) 139.
\bibitem{Lucha98R}W.~Lucha and F.~F.~Sch\"oberl, in: Proceedings of the
XI$^{\rm th}$ International Conference ``Problems of Quantum Field Theory,''
edited by B.~M.~Barbashov, G.~V.~Efimov, and A.~V.~Efremov, July 13 -- 17,
1998, Dubna, Russia (Joint Institute for Nuclear Research, Dubna, 1999),
p.~482.
\bibitem{LTS}S.~Jacobs, M.~G.~Olsson, and C.~Suchyta III, Phys.~Rev.~D {\bf
33} (1986) 3338; {\bf 34} (1986) 3536 (E).
\bibitem{Abramowitz}{\em Handbook of Mathematical Functions}, edited by
M.~Abramowitz and I.~A.~Stegun (Dover, New York, 1964).
\bibitem{Falkenst85}P. Falkensteiner, H. Grosse, F. Sch\"oberl, and P.
Hertel, Comput. Phys. Comm. {\bf 34} (1985) 287; for the Mathematica 3.0
update of this routine, see: W.~Lucha~and F.~F.~Sch\"oberl, 
Int. J. Mod. Phys. C {\bf 10} (1999) 607.
\end{thebibliography}
\end{document}